\documentclass[aps,draft]{revtex4}
\usepackage[dvips]{graphicx}
\begin{document}

\title{New Longitudinal Waves in Electron-Positron-Ion Quantum Plasmas}

\author{ Nodar L. Tsintsadze$^{1,2}$, Levan N. Tsintsadze$^{2,3}$, A. Rehman$^1$, G. Murtaza$^1$}
\affiliation{ $^1$ Salam Chair in Physics, GC University, Lahore-54000, Pakistan \\ $^2$ Department of Plasma Physics, E.Andronikashvili Institute of Physics, Tbilisi, Georgia \\ $^3$ Graduate School of Science, Hiroshima University, Higashi-Hiroshima, Japan}

\date{\today}

\begin{abstract}

A general quantum dispersion equation for electron-positron(hole)-ion quantum plasmas is derived and studied for some interesting cases. In an electron-positron degenerate Fermi gas, with or without the Madelung term, a new type of zero sound waves are found. Whereas in an electron-hole plasmas a new longitudinal quantum waves are revealed, which have no analogies in quantum electron-ion plasmas. The excitation of these quantum waves by a low-density monoenergetic straight electron beam is examined. Furthermore, the KdV equation for novel quantum waves is derived and the contribution of the Madelung term in the formation of the KdV solitons is discussed.

\end{abstract}

\pacs{52.35.-g, 52.27.-h}

\maketitle

\section{Introduction}

In recent years, a huge number of works have been devoted to
the investigation of the collective behavior of quantum plasmas (for review see \cite{pad}).
Such interest is motivated by its potential application in modern
technology, e.g. metallic and semiconductor nanostructures - such as metallic nanoparticles, metal clusters, thin metal films, spintronics, nanotubes, quantum well and quantum dots, nano-plasmonic devices, quantum x-ray free-electron lasers, etc. Moreover, quantum electron-ion or electron-positron-ion plasmas are common in planetary interiors, in compact astrophysical objects (e.g., the interior of white dwarf stars, magnetospheres of neutron stars and magnetars, etc.), as well as in the next generation intense laser-solid density plasma experiments. The quantum plasma may arise when a pellet of hydrogen is compressed to many times the solid density in the fast ignition scenario for inertial confinement fusion.

In the past the properties of linear electron oscillations in a dense Fermi plasma have been studied in Refs.\cite{gol}-\cite{bp}. A new type of quantum kinetic equations of the Fermi particles of various species were derived recently \cite{tsin}, and a general set of fluids equations describing the quantum plasma was obtained. This kinetic equation for the Fermi quantum plasma was used in Ref.\cite{tsin} to study the propagation of small longitudinal perturbations in an electron-ion collisionless plasmas, deriving a quantum dispersion equation. The dispersion properties of electrostatic oscillations in quantum plasmas have been discussed later in Refs.\cite{tsin10}, \cite{ell}. The effects of the quantization of the orbital motion of electrons and the spin of electrons on the propagation of longitudinal waves in the quantum plasma have been also reported \cite{ltsin}.
Instabilities of the quantized longitudinal electric waves \cite{ltsin} due to an electron beams were investigated very recently in \cite{ltsin10}.

In the present paper, we consider two different quantum plasmas:
one is the Fermi gas composed of electrons, positrons and ions, 
and the other is of electrons, holes and ions. Note that the electrons and holes have different
mass due to interaction between particles. For instance, in a solid-state
plasma or a quantum liquid the effective mass of charge
carriers (electrons and holes) differs from that of free electrons.
Semiconductors, containing light negative (electrons) and heavy
positive (holes) charge carriers, can be degenerate ($n_e\geq
10^{16}-10^{18}\ cm^{-3}$) with the effective mass of electrons
$m^\ast_e\approx(0.01-0.1)m_e$, and the degeneracy occurs at
temperatures $T<10^2 K$.
In the degenerate solid-state plasma or the Fermi liquid, although the
particles are crowded together only a few $A^o$ apart, the mean free path
is longer than a few centimeters at low temperature. Two factors are
responsible for such long mean free paths. One is the Pauli
exclusion principle that allows collisions only to final states,
which were vacant before the collisions. The second factor is the
screening of the Coulomb interaction between the particles \cite{lan}-\cite{tsin09}.
Hence, in our investigation we can suppose the plasma to be collisionless
at low temperatures, and we take into account the linear Landau
damping. Unless otherwise stated, the ions are assumed to be immobile in this paper.

\section{Basic Equations}

Considering the propagation of small longitudinal perturbations (${\bf H}=0\ ,{\bf E}=-\nabla \phi$) in an electron-positron(hole)-ion plasmas, the relevant quantum kinetic equation is as follows \cite{tsin}
\begin{eqnarray}
\label{qke}
\frac{\partial f_\alpha}{\partial t}+({\bf v}\cdot\nabla)f_\alpha-
e_\alpha\nabla \phi\frac{\partial f_\alpha}{\partial \bf
p}+\frac{\hbar^2}{2m_\alpha}\nabla\frac{1}{\sqrt{n_\alpha}}\triangle\sqrt{n_\alpha}\:\frac{\partial
f_\alpha}{\partial \bf p}= 0 \ ,
\end{eqnarray}
where the last term is the Madelung term, the suffix $\alpha$ stands for the particle species, $\hbar$ is the Plank constant divided by $2\pi$ and the other notation is standard.
We
express a distribution function as $f_\alpha=f_{\alpha
0}+\delta f_\alpha$, where $f_{\alpha 0}$ is the unperturbed,
stationary isotropic homogeneous distribution function and $\delta
f_\alpha$ is the small perturbed part. Linearizing Eq.(\ref{qke}), supposing
that $\delta f_\alpha$ and $\delta \phi$ vary like $exp\imath({\bf
k}\cdot{\bf r}-\omega t)$, and combining it with the Poisson equation, we
obtain the following general quantum dispersion relation
\begin{eqnarray}
\label{gd}
1+\Sigma_{\alpha}\:\frac{3\omega^2_{p\alpha}}{k^2
v^2_{F\alpha}}\frac{1}{\Gamma_\alpha}\left[1-\frac{\omega}{2kv_{F\alpha}}\:ln\frac{\omega+k
v_{F\alpha}}{\omega-k v_{F\alpha}}\right]=0\ ,
\end{eqnarray}
where
\begin{eqnarray}
\label{ga}
\Gamma_\alpha=1+\frac{3\hbar^2 k^2}{4 m^2_\alpha v^2_{F\alpha}
}\left(1-\frac{\omega}{2kv_{F\alpha}}\:ln\frac{\omega+k
v_{F\alpha}}{\omega-k v_{F\alpha}}\right).
\end{eqnarray}
In obtaining Eq.(\ref{gd}) we have assumed $T_F>>T$ (where $T_F$ is the
Fermi degeneracy temperature and $T$ is the thermal temperature), and
the Fermi distribution function to be the step
function $f_{\alpha 0}=\Theta(\varepsilon_{F\alpha}-\varepsilon)$, where $\varepsilon_{F\alpha}=m
v^2_{F\alpha}/2=\frac{(3\pi^2)^{2/3}\hbar^2}{2m_\alpha}n_\alpha^{2/3}$. In the following we examine the dispersion Eq.(\ref{gd}), which describes
oscillation properties both of electrons and positrons, for some interesting cases.

\section{Zero Sound Waves}

We first consider the zero sound waves in an electron-positron-ion gas without the Madelung term. From
Eq.(\ref{gd}), choosing the frequency range $k v_{Fe} > \omega \sim k
v_{Fp}\:$, and taking $m_e=m_p$ we get the real and imaginary part of frequency as
\begin{eqnarray}
\label{rf}
\omega_r=kv_{Fp}\left[1-2
\:exp\left(-2\left[1+k^2\lambda^2_p+\left(\frac{n_{oe}}{n_{op}}\right)^{1/3}\right]\right)\right]
\end{eqnarray}
and
\begin{eqnarray}
\label{if}
\omega_i=-4\pi k
v_{Fp}\:exp\left(-4\left[1+k^2\lambda^2_p+\left(\frac{n_{oe}}{n_{op}}\right)^{1/3}\right]\right)\ ,
\end{eqnarray}
where $\lambda_p=v_{Fp}/\sqrt{3}\:\omega_{pp}$ is the positron
Thomas-Fermi screening length. We notice that the exponential
damping in the imaginary part of frequency is twice stronger than in the real one. It should be noted that in an electron-ion plasma zero sound waves don't have the imaginary part of frequency. However, in the case considered here, it does exist owing to positrons. The physical reason for such damping is obvious from Eq.(\ref{if}).

Under the same assumptions as above, the frequencies of zero sound waves with the Madelung term
are modified as
\begin{eqnarray}
\label{mrf}
\omega_r=k\:v_{Fp}\left[1+2\:exp\left(-2\left[1+\frac{k^2 \lambda^2_p
+\left(\frac{n_{oe}}{n_{op}}\right)^{1/3}}{1+a}\right]\right)\cos\left(\frac{\pi}{1+a}\right)\right]
\end{eqnarray}
\begin{eqnarray}
\label{mif}
\omega_i=-2k\:v_{Fp}\:exp\left(-2\left[1+\frac{k^2 \lambda^2_p
+\left(\frac{n_{oe}}{n_{op}}\right)^{1/3}}{1+a}\right]\right)
\sin\left(\frac{\pi}{1+a}\right)\ ,
\end{eqnarray}
where $\:\:a=\frac{3\:\omega^2_q}{k^2
v^2_{Fp}}\left(\frac{n_{oe}}{n_{op}}\right)^{1/3}$ can take any value, which plays an important role, and $\ \omega_q=\frac{\hbar k^2}{2 m_e}\ $ is the quantum
oscillation frequency. For example, in the case a=1, one gets purely quantum effect
\begin{eqnarray*}
\omega_r\simeq k v_{Fp} \hspace{1cm} and \hspace{1cm} \omega_i=-2 k v_{Fe} exp\left[-\left(k^2 \lambda^2_p
+(\frac{n_{oe}}{n_{op}})^{1/3}\right)\right]\ .
\end{eqnarray*}
It should be noted that in
usual zero sound waves one has $k^2 \lambda^2_p>>1$,
but here the zero sound waves exist even in the limit $k^2 \lambda^2_p\approx1$ due to the factor $(n_{oe}/n_{op})^{1/3}$ in the exponential part of Eqs.(\ref{rf}-\ref{mif}).

\section{Special Cases with $m_p>m_e$}

We now study special cases in which the effective mass of electron
is less than the mass of positron. In semiconductors, for
example, we often have such situation when the mass of the hole becomes
much greater than the effective mass of the electron.

First, we consider the case without the Madelung term, supposing that the ions are stationary.
For the frequency range $k v_{Fe} > \omega > k v_{Fp}$, we can write Eq.(\ref{gd}) as
\begin{eqnarray}
\label{sco}
1+\frac{3\:\omega^2_{pe}}{k^2 v^2_{Fe}}\left(1+\frac{i
\pi\:\omega}{2 k
v_{Fe}}\right)-\frac{\omega^2_{pp}}{\omega^2}=0\ .
\end{eqnarray}
From Eq.(\ref{sco}), separating the real and the imaginary parts of the frequency, it
is straightforward to obtain the following relations
\begin{eqnarray}
\label{sr}
\omega_r=\:\frac{1}{\sqrt{3}}\left(\frac{m_e}{m_p}\frac{n_{op}}{n_{oe}}\right)^{1/2}k
v_{Fe}\ ,
\end{eqnarray}
\begin{eqnarray}
\label{si}
\omega_i=-\frac{\pi}{12}\left(\frac{m_e}{m_p}\frac{n_{op}}{n_{oe}}\right)k
v_{Fe}\ .
\end{eqnarray}
We should emphasize that the real part of the frequency is larger than the imaginary one, since
$\left(\frac{m_e}{m_p}\frac{n_{op}}{n_{oe}}\right)^{1/2}<<1$. We also note that Eqs.(\ref{sr}) and (\ref{si}) describe the propagation of new waves with slow damping.

Next, in the case when the Madelung term is incorporated, the equation (\ref{gd}) takes the form
\begin{eqnarray}
\label{sm}
1+\frac{3\omega^2_{pe}}{k^2
v^2_{Fe}}\frac{1}{\Gamma_e}\left(1+\frac{i\pi \omega}{2 k
v_{Fe}}\right)-\frac{\omega^2_{pp}}{\omega^2}\frac{1}{\Gamma_p}=0\ ,
\end{eqnarray}
where
\begin{eqnarray*}
\Gamma_e=1+\frac{3\omega^2_{qe}}{k^2
v^2_{Fe}}\left(1+\frac{i\pi \omega}{2 k v_{Fe}}\right)
\end{eqnarray*}
and
\begin{eqnarray*}
\Gamma_p=1-\frac{\omega^2_{qp}}{\omega^2}\ .
\end{eqnarray*}

From Eq.(\ref{sm}), the real frequency can be expressed as
\begin{eqnarray}
\label{crf}
\omega^2_r=\omega^2_{qp}+\frac{\omega^2_{pp}}{1+\frac{3\:\omega^2_{pe}}{k^2
v^2_{Fe}}} \ ,
\end{eqnarray}
which for $k^2v^2_{Fe}>>\omega^2_{pe}$ and
$k^2v^2_{Fe}<<\omega^2_{pe}\:$, reads respectively
\begin{eqnarray}
\label{cfrf}
\omega^2_r=\omega^2_{qp}+\omega^2_{pp}
\end{eqnarray}
and
\begin{eqnarray}
\label{frf}
\omega^2_r=\omega^2_{qp}+\left(\frac{m_e}{m_p}\frac{n_{op}}{n_{oe}}\right)\frac{k^2
v^2_{Fe}}{3}\ .
\end{eqnarray}
The corresponding damping rates are given by
\begin{eqnarray*}
\omega_i=-\frac{3\pi}{4}\:\omega_{pp}\frac{\omega_{pp}}{kv_{Fe}}\left(\frac{\omega_{pe}}{kv_{Fe}}\right)^2\ ,
\end{eqnarray*}
\begin{eqnarray*}
\omega_i=-\frac{\pi}{12}\left(\frac{m_e}{m_p}\frac{n_{op}}{n_{oe}}\right)k
v_{Fe}\ .
\end{eqnarray*}
Note here that the quantum (Madelung) term does not participate in the damping process. It should be also emphasized that new modes found in this section, exist only when $m_p>m_e$.

\section{Beam Plasma Interaction}

We now propose the excitation of the
new type of waves (\ref{cfrf}) and (\ref{frf}) by a straight electron beam, with
the density $n_b$ much less than the plasma density, which is
injected into a degenerate electron-hole plasma. We assume that the
electron beam obeys the  Maxwellian distribution function (since the
density is low). Starting from Eq.(\ref{qke}), and following the usual
procedure, we obtain the dispersion relation including the electron beam
contribution
\begin{eqnarray}
\label{deb}
1+\delta\epsilon_e+\delta\epsilon_p+\delta\epsilon_b=0\ ,
\end{eqnarray}
where
\begin{eqnarray}
\label{wh}
\delta\epsilon_\alpha=\frac{3\omega^2_{p\alpha}}{k^2
v^2_{F\alpha}}\frac{1}{\Gamma_\alpha}\left[1-\frac{\omega}{2kv_{F\alpha}}\:ln\frac{\omega+k
v_{F\alpha}}{\omega-k
v_{F\alpha}}\right]\ , \hspace{1cm} \alpha=e,p
\end{eqnarray}
and
\begin{eqnarray}
\label{whb}
\delta\epsilon_b=\frac{\omega^2_b}{k^2
v^2_{tb}}\frac{1}{\Gamma_b}\left[1-I_+\left(\frac{\omega^\prime}{k\:v_{tb}}\right)\right]\ .
\end{eqnarray}
Here $\:\omega^\prime=\omega-{\bf k}\cdot{\bf u}\:$,
${\bf u}$ is the velocity of electron beam,
$\ I_+(x)=xe^{-x^2/2}\int_{i\infty}^x\:d\tau\:e^{\tau^2/2}\ $,
and
\begin{eqnarray*}
\Gamma_b = 1+\frac{\omega^2_{qb}}{k^2 v^2_{tb}}\left[1-I_+\left(\frac{\omega^\prime}{k\:v_{tb}}\right)\right]\ .
\end{eqnarray*}

In the frequency range $k v_{Fe} > \omega > k v_{Fp}$
using the asymptotic value \cite{ale} of the function
$I_+(\frac{\omega^\prime}{k\:v_{tb}})$ for $\omega^\prime
>>k\:v_{tb}$, from Eq.(\ref{deb}) we get
\begin{eqnarray}
\label{fdb}
1+\frac{3\:\omega^2_{pe}}{k^2 v^2_{Fe}}\left(1+\frac{i
\pi\:\omega}{2 k
v_{Fe}}\right)-\frac{\omega^2_{pp}}{\omega^2-\omega^2_{qp}}-\frac{\omega^2_{pb}}{(\omega-{\bf
k}\cdot{\bf u})^2-\omega^2_{qb}}=0\ .
\end{eqnarray}

Examining the dispersion relation (\ref{fdb}), we consider the hydrodynamic
instability in the long wavelength limit, $3\omega^2_{pe}>>k^2
v^2_{Fe}$. For $\omega=\omega_r+\gamma$ (where $\omega_r$ is given by Eq.(\ref{frf})) and $\omega={\bf
k}\cdot{\bf u}-\omega_{qb}+\gamma$, with $|\gamma|<<\omega\:,\:$  Eq.(\ref{fdb}) yields
\begin{eqnarray}
\label{im}
Im\:\gamma=\frac{1}{2}\left(\frac{m_p}{m_b}\frac{n_{ob}}{n_{op}}\right)^{1/2}
\frac{\omega^2_r-\omega^2_{qp}}{\sqrt{\omega_r\:\omega_{qb}}}
\end{eqnarray}
or
\begin{eqnarray}
\label{imo}
Im\:\gamma=\frac{1}{6}\frac{\sqrt{n_{ob}\:n_{op}}}{n_{oe}}\frac{m_e}{\sqrt{m_b\:m_p}}
\frac{k^2v^2_{Fe}}{\sqrt{\omega_r\:\omega_{qb}}}\ .
\end{eqnarray}
It should be noted that this growth rate is also purely quantum.

For the same frequency range $k v_{Fe} >
\omega > k v_{Fp}$, but with $\omega^\prime <<k\:v_{tb}\:$, Eq.(\ref{deb}) casts in the form
\begin{eqnarray}
\label{ki}
1+\frac{3\:\omega^2_{pe}}{k^2 v^2_{Fe}}\left(1+\frac{i
\pi\:\omega}{2 k
v_{Fe}}\right)-\frac{\omega^2_{pp}}{\omega^2-\omega^2_{qp}}+\frac{\omega^2_{pb}}{k^2
v^2_{tb}}
\left(1+i\sqrt{\frac{\pi}{2}}\frac{\omega^\prime}{kv_{tb}}\right)=0\ ,
\end{eqnarray}
which leads to the following result
\begin{eqnarray}
\label{kio}
\omega_i=-\frac{3\pi}{4}\left(\frac{m_p}{m_e}\frac{n_{oe}}{n_{op}}\right)\frac{(\omega^2_r-\omega^2_{qp})^2}{k^3v^3_{Fe}}
-\frac{1}{2}\sqrt{\frac{\pi}{2}}\left(\frac{m_p}{m_b}\frac{n_{ob}}{n_{op}}\right)\frac{\omega_r-{\bf
k}\cdot{\bf
u}}{\omega_r\:k^3v^3_{tb}}\:(\omega^2_r-\omega^2_{qp})^2\ .
\end{eqnarray}
So that for the kinetic instability the following inequality
\begin{eqnarray*}
{\bf k}\cdot{\bf
u}>\left(1+3\:\sqrt{\frac{\pi}{2}}\:\frac{m_b}{m_e}\:\frac{n_{oe}}{n_{ob}}\:\frac{v^3_{tb}}{v^3_{Fe}}\right)\omega_r
\end{eqnarray*}
should be satisfied.

\section{KdV-Equation from Quantum Hydrodynamic Model}

In order to discuss the nonlinear effects in the electron-positron(hole)-ion plasmas in the hydrodynamic approximation, following the standard procedure \cite{lanf},\cite{kar} we now derive KdV equation, recalling that the positrons (holes) are heavier than the electrons ($m_p>m_e$) and $n_{oe}>n_{op}$. For our purpose, we employ the hydrodynamic equations derived in Ref.\cite{tsin}
\begin{eqnarray}
\label{pem}
\frac{\partial{\bf v}_\alpha}{\partial t}+({\bf
v}_\alpha\cdot\nabla){{\bf v}_\alpha}=-\frac{e_\alpha\nabla
\phi}{m_\alpha}-\frac{1}{m_\alpha n_\alpha}\nabla
P_\alpha+\frac{\hbar^2}{2m^2_\alpha}\nabla\frac{1}{\sqrt{n_\alpha}}\triangle\sqrt{n_\alpha}\ ,
\end{eqnarray}
\begin{eqnarray}
\label{pec}
\frac{\partial n_\alpha}{\partial t}+\nabla(n_\alpha {\bf v}_\alpha)=0\ ,
\end{eqnarray}
and the Poisson equation reads
\begin{eqnarray}
\label{pp}
\nabla^2\phi=4\pi en_{oe}(N_e-N_p-N_{oi})\ ,
\end{eqnarray}
where $P_\alpha=\frac{(3\pi^2)^{2/3}\hbar^2}{5m_\alpha}n_\alpha^{5/3}$, $\ N_e=n_e/n_{oe}\ $, $\ N_p=n_p/n_{oe}\ $, and $\ N_{oi}=n_{oi}/n_{oe}\ $. Note that the pressure of electrons is larger than the pressure of positrons, since  $n_{oe}>n_{op}$ and $m_e<m_p$.

Hereafter, we shall consider only low frequency electron-positron
oscillations, the dispersion relation of which is given by Eq.(\ref{crf}).
The condition $\omega<<k v_{Fe}\:$, allows us to consider electrons as inertialess. Combining equations of motion (\ref{pem}) for electrons and positrons, and taking into account $P_e>P_p$, in one dimensional case we obtain
\begin{eqnarray}
\label{od}
\frac{\partial v_p}{\partial t}+v_p\ \frac{\partial v_p}{\partial x}=-\frac{\varepsilon_{Fe}}{m_p}\frac{\partial}{
\partial x}\ N_e^{2/3}+\frac{\hbar^2}{2m_em_p}\ \frac{\partial}{\partial x}\ \frac{1}{\sqrt{N_e}}\ \frac{\partial^2}{\partial x^2}\ \sqrt{N_e} \ .
\end{eqnarray}

In order to construct the KdV equation, as in classical plasmas, here we must also assume that the second derivative of the potential field is much less than each term on the right-hand side (R.H.S.) in Eq.(\ref{pp}). Defining $e\phi$ from the equation of motion for electrons and substituting it in the Poisson equation, we get
\begin{eqnarray}
\label{ne}
N_e=N_p+N_{oi}+\lambda_e^2\ \frac{\partial^2}{\partial x^2}\ N_e^{2/3} \ ,
\end{eqnarray}
where $\lambda_e=\frac{1}{\sqrt{3}}\frac{v_{Fe}}{\omega_{pe}}$ is the electron Thomas-Fermi screening length.

Use of Eq.(\ref{ne}) in Eq.(\ref{od}) yields the relation
\begin{eqnarray}
\label{use}
\frac{\partial v_p}{\partial t}+v_p\frac{\partial v_p}{\partial x}=-\frac{v_s^2}{2}\frac{\partial}{
\partial x}(N_p+N_{oi})^{2/3}-\frac{v_s^2}{3}\lambda_e^2\frac{\partial^3}{\partial x^3}(N_p+N_{oi})^{2/3} +\frac{\hbar^2}{2m_em_p}\frac{\partial}{\partial x}\frac{1}{\sqrt{N_p+N_{oi}}}\frac{\partial^2}{\partial x^2} \sqrt{N_p+N_{oi}},
\end{eqnarray}
where $v_s=\sqrt{\frac{2\varepsilon_{Fe}}{m_p}}$.

We must also find the relation between $v_p$ and $N_p$. To this end, we use a simple wave solution \cite{lanf},\cite{kar}, which means that $v_p$ is the function of the density $n_p$ alone (i.e., $v_p(n_p)$ or $n_p(v_p)$), for the system of Eq.(\ref{use}) and the equation of continuity of positrons. The result is
\begin{eqnarray}
\label{int}
\frac{\sqrt{3}}{v_s}v_p=\int\frac{d N_p}{\sqrt{N_p
(N_p+N_{oi})^{1/3}}}\ .
\end{eqnarray}

To derive an explicit relation between $v_p$ and $N_p$, we now consider the integral (\ref{int}) in two limiting cases. First, for the case $N_{oi}>N_p$, from Eq.(\ref{int}) we have
\begin{eqnarray}
\label{vfc}
N_p=N_{op}\left(1+\frac{\sqrt{3}}{2}\frac{N_{oi}^{1/6}}{\sqrt{N_{op}}}\frac{v_p}{v_s}\right)^2 \ .
\end{eqnarray}
Substituting this expression into Eq.(\ref{use}), and replacing $v_p$ by $u=\frac{3}{2}v_p$, we obtain
\begin{eqnarray}
\label{vfcr}
\frac{\partial u}{\partial t}+(v_1+u)\frac{\partial u}{\partial x}=\frac{\sqrt{3}}{v_s}\frac{N_{op}^{1/2}}{N_{oi}^{5/6}}\left(\frac{\hbar^2}{4m_em_p}-\frac{v_s^2\lambda_e^2}{3}
N_{oi}^{2/3}\right)\frac{\partial^3 u}{\partial x^3}\ ,
\end{eqnarray}
where $v_1=\frac{1}{2}\frac{N_{op}^{1/2}}{N_{oi}}v_s$. Note that R.H.S. of Eq.(\ref{vfcr}) can be positive or negative depending on the plasma parameters.

Next in the case, when $N_p>N_{oi}$, after a simple integration of Eq.(\ref{int}) we get
\begin{eqnarray}
\label{vf}
N_p=N_{op}\left(1+\frac{1}{\sqrt{3}N_{op}^{1/3}}\frac{v_p}{v_s}\right)^3 \ .
\end{eqnarray}
Use of the expression (\ref{vf}) in Eq.(\ref{use}) yields
\begin{eqnarray}
\label{vfy}
\frac{\partial V}{\partial t}+(v_2+V)\frac{\partial V}{\partial x}=\frac{\sqrt{3}}{v_s N_{op}^{1/3}}\left(\frac{\hbar^2}{4m_em_p}-\frac{v_s^2\lambda_e^2}{3}
N_{op}\right)\frac{\partial^3 V}{\partial x^3}\ ,
\end{eqnarray}
where $v_2=\frac{4N_{op}^{1/3}}{3\sqrt{3}} v_s$ and $V=\frac{4}{3}v_p$.

We emphasize here that Eqs.(\ref{vfcr}) and (\ref{vfy}) have two physically distinct solutions. Namely, if the R.H.S. of these equations are negative, then the solutions of Eqs.(\ref{vfcr}) and (\ref{vfy}) correspond to the compressional solitons. Whereas, in the opposite case, when the quantum (Madelung) term exceeds the term due to the charge separation, the solitary wave is a rarefaction wave.

\section{Summary}

We have investigated the propagation of small longitudinal perturbations in an electron-positron-ion and electron-hole-ion plasmas, deriving a general quantum dispersion equation. Studying this dispersion relation, in the electron-positron degenerate Fermi gas, with or
without the Madelung term, we have revealed a new type of zero sound waves,
which slowly damp in the range of frequencies $\omega\leq kv_{Fp}$. Whereas, in an electron-hole plasmas we found a new longitudinal quantum waves, which have no analogies in quantum electron-ion plasmas. We have proposed the generation of these new waves by a straight electron beam, with the density much less than the plasma density. Finally, we have derived KdV equation in order to discuss the nonlinear effects in electron-positron(hole)-ion plasmas, and found that due to the quantum (Madelung) term the compressional solitons may become rarefaction waves. These investigations may play an essential role for the description of complex phenomena that appear in dense astrophysical objects, and in the next generation intense laser-solid density plasma experiments, as well as may have a potential application in modern technology.

\end{document}